\documentclass
[preprint,showpacs,byrevtex,10pt,a4paper,twocolumn,tightenlines]{revtex4}%
\usepackage{amsfonts}
\usepackage{amsmath}
\usepackage{amssymb}
\usepackage[dvips]{graphicx}%
\begin{document}
\preprint{ }
\title{Optical excitation of Electron-Glasses}
\author{Z. Ovadyahu}
\affiliation{Racah Institute of Physics, The Hebrew University, Jerusalem 91904, Israel }

\begin{abstract}
Electron-glasses can be readily driven far from equilibrium by a variety of
means. Several mechanisms to excite the system and their relative merits are
reviewed. In this study we focus on the process of exciting electron-glasses
by interaction with near infrared radiation. The efficiency of this protocol
varies considerably among different electron-glasses, but it only weakly
depends on their resistance at liquid helium temperatures. A dramatic
enhancement of the excitation efficiency is observed upon doping crystalline
indium-oxide with Au. Some enhancement is observed also in samples doped with
Pb but this enhancement fades away with time unlike the situation in the
Au-doped samples. Several structural and analytical tools are used to
characterize the changes in the materials that may be responsible for these
effects. Possible routes by which high-frequency electromagnetic fields take
the system far from equilibrium are discussed.

\end{abstract}
\maketitle

\section{Introduction}

Non-equilibrium transport features, expected to occur in electron-glasses
\cite{1,2,3,4,5}, have been reported in several systems over the last two
decades \cite{6,7,8,9,10,11}. A common way to excite the system to observe
these phenomena has been a sudden change of the gate voltage using samples
configured as three terminal devices \cite{6,7,8,9,11}. This gating technique
is straightforward to apply and may yield very useful information on the
nature of the glassy state (such as the memory-dip \cite{12}). However, it may
not be possible for technical reasons to configure the sample with a gate, and
in a bulk sample this technique is ineffective. It may also give rise to
artifacts associated with trapped charge or other defects in the insulating
layer that separates the sample from the gate (this appears to be a common
problem in some semiconductors). It is therefore desirable to find different
ways to take the electron-glass far from equilibrium, and in fact there are
several alternatives to achieve that.

A readily available way to excite the electron-glass is to employ a non-ohmic
field F along the sample (a longitudinal field) \cite{10,13}. In this protocol
the system is taken from equilibrium to a degree that depends on the applied F
as well as on the time the sample is under the stress of the field.
Qualitatively similar effect may be obtained by increasing the bath
temperature T by $\Delta$T and maintaining it for a while \cite{13,14}. The
$\Delta$T-protocol is experimentally harder to control and it is usually
difficult to quickly change and maintain the bath temperature. This is where
the F-protocol has a distinct advantage over the $\Delta$T-protocol but the
underlying mechanism for excitation seems to be essentially the same;
increasing the population of high-energy phonons.

It is natural to expect that the electron-glass could be also excited by
exposure to electromagnetic (EM) radiation. It turns out that this depends on
the frequency $\omega$ of the EM field. Efficient excitation has been observed
in experiments on indium-oxide samples where a brief exposure to an infra-red
(IR) radiation led to a slowly relaxing conductance component \cite{15}. Also,
it was demonstrated that exposure to IR source effectively erases the
memory-dip \cite{15}. Exposing the sample to EM radiation at the microwaves
frequencies (tested in the range 1-100GHz) on the other hand, was not
effective in this regard, in particular, it has no effect on the memory-dip
\cite{13,16}. As long as the microwaves field was on, it did increase the
conductance by as much (and even more) as the IR acting on the same sample but
once the microwave source was turned off the conductance returned quickly to
its equilibrium value.

These empirical results are consistent with the following physical picture.
The system is in equilibrium when the occupation of the electronic states,
under a given potential landscape and temperature, minimizes the free energy.
In the interacting system this leads to a specific organization of the way the
localized states are occupied. In the electron-glass it is the Coulomb
interaction that introduces correlations between states-occupation and their
spatial coordinates. Among other things this process leads to a Coulomb gap in
the single-particle density of states \cite{17}. Randomizing such a
configuration requires an energy investment that exceeds E$_{C}$,$~$which is
of the order of the Coulomb energy \cite{16}. Re-establishing the equilibrium
configuration from a state where the sites are randomly distributed is a slow
process for reasons discussed elsewhere \cite{16}. Therefore, $\Delta$G(t)
produced by such an excitation agent will decay slowly.

Excitation by the $\Delta$T-protocol (or the F-protocol) differs from other
protocols in that it usually involves long application-times to be effective.
Note that only high-frequency phonons (with energy quantum
%TCIMACRO{\TEXTsymbol{>}}%
%BeginExpansion
$>$%
%EndExpansion
E$_{C}$) can do the job, and such energetic modes are exponentially rare when
$\Delta$T$\ll$E$_{C}$. Therefore to achieve a degree of randomization that
results in a slowly relaxing $\Delta$G component in the $\Delta$T-protocol (or
the F-protocol) requires a long time of applying the excitation agent
(depending on $\Delta$T). By contrast, brief exposure to IR is sufficient to
effectively randomize the system as will be demonstrated in section III. Given
the ease and flexibility of applying the IR-protocol, one wonders why it has
been applied so sparingly in the study of electron-glasses \cite{18}.

In this work we compare the efficiency of this protocol among different
electron-glasses; amorphous and crystalline indium-oxide (In$_{x}$O and
In$_{2}$O$_{3-x}$ respectively), granular-aluminum, and beryllium. The main
focus of this work is on the dramatic change of the excitation efficiency of
In$_{2}$O$_{3-x}$ samples when doped with Au and Pb. Possible reasons for this
enhancement are experimentally tested and the implications to the physics
underlying the IR-protocol are discussed.

\section{Experimental}

\subsection{Sample preparation and measurement techniques}

Several materials were used to prepare samples for measurements in this study.
These were thin films of crystalline or amorphous indium-oxide (In$_{2}%
$O$_{3-x}$ and In$_{x}$O respectively), and granular-aluminum. These three
materials were similarly prepared by e-gun evaporation of In$_{2}$O$_{3}$ or
aluminum pellets unto room temperature glass substrates in partial oxygen
pressures of 4-6$\cdot$10$^{-5}$mbar and rates of 0.3-1\AA /second. The
In$_{2}$O$_{3-x}$ samples were obtained from the as-deposited amorphous films
by crystallization at 525K (unless otherwise mentioned). Also used in this
work are films of In$_{2}$O$_{3-x}$ doped with either Au or Pb. These dopants
were deposited from a Knudsen source simultaneously with the indium-oxide, and
the composite deposit was then crystallized in the same manner as the In$_{2}%
$O$_{3-x}$. The relative concentration of a dopant was controlled in the
deposition process by independently measuring the evaporation rate of each
material. Each preparation run produced a batch of samples on a common
glass-slide. Each batch included samples for excitation measurements, samples
for Hall-effect measurements, and samples for structural and chemical
analysis. Carbon coated Cu grids were attached to the substrate for the
electron microscopy study.

X-ray photoelectron-spectroscopy (XPS, using Axis Ultra) and
transmission-electron-microscopy (TEM, using Tecnai F20 G2) were employed to
characterize the films composition and microstructure. Optical studies
employed the Cary-1 spectrophotometer and Bruker Equinox-55 FTIR.

Indium contacts were employed in all indium-oxide based samples. To make good
quality contacts to the granular-aluminum films gold strips were deposited on
the ion-bombarded glass substrates prior to the aluminum deposition. Lateral
size of the samples used here were 0.2-1mm. Their thickness (typically,
40-55\AA ~for In$_{2}$O$_{3-x},$In$_{2}$O$_{3-x}$:Pb, In$_{2}$O$_{3-x}$:Au,
150-600\AA ~for In$_{x}$O, and 50-60\AA ~for granular-aluminum) and, for
In$_{2}$O$_{3-x}$ based samples, stoichiometry, were chosen such that at the
measurement temperatures (at or close to 4K), the samples had sheet resistance
R$_{\square}$ in the range 0.2M$\Omega$-3G$\Omega$. This ensured that all
these samples are deep enough in the hopping regime to exhibit electron-glass
features. All resistance data reported below are per square geometry. We also
show for comparison results on beryllium films using a specimen studied in
\cite{11}.

Optical excitations in this work were accomplished by exposing the sample to
an AlGaAs diode (operating at $\approx$0.88$\pm$0.05$\mu$m), placed typically
$\approx$10-25mm from the sample. The diode was energized by a
computer-controlled Keithley 220 current-source.

Conductivity of the samples was measured using a two terminal ac technique
employing a 1211-ITHACO current preamplifier and a PAR-124A lock-in amplifier
for the 4K studies. All measurements reported here were performed with the
samples immersed in liquid helium at T$\approx$4.1K held by a 100 liters
storage-dewar. The ac voltage bias was small enough to ensure near-ohmic
conditions (except when it was desirable to simulate higher temperature
behavior). Complementary details of sample preparation, characterization, and
measurements techniques are given elsewhere \cite{6,19}.

\section{Results and discussion}

\subsection{The basic facts}

The experimental protocol used for the IR excitation is illustrated in Figure
1 using a rather thick In$_{x}$O sample. Starting by monitoring G(t) for
several minutes to establish a baseline that defines the near-equilibrium
conductance G$_{0}$. The IR source is then turned on for 3 seconds, and then
turned off while G(t) continues to be measured. The brief IR burst causes G to
promptly increase by $\Delta$G some of which quickly decays once the source is
turned off leaving a slow relaxation tail as shown in figure 1. The inset to
the figure demonstrates that the slow component of the induced excess
conductance relaxes logarithmically with time, which is a characteristic
feature of the electron-glass \cite{20}. It also depicts the way its `initial'
amplitude $\delta$G is defined in this work (i.e., by extrapolating the
logarithmic G(t) to t=1 second, the typical resolution-time in our
experiments). The fast component of the relaxation, $\Delta$G-$\delta$G, is
essentially a recovery from the thermal shock associated with the IR burst.
%TCIMACRO{\FRAME{ftbpFU}{3.1012in}{3.1012in}{0pt}{\Qcb{IR excitation protocol
%applied on a 600\AA ~In$_{x}$O film and R$_{\square}$=50M$\Omega.$ The
%parameters that characterize the results are marked by arrows; $\Delta$G is
%the conductance initial jump, $\delta$G is the initial amplitude for the
%slowly relaxing part which is defined in this work as the value of the excess
%conductance at t=t$_{S}$ which is the time corresponding to t=1s of the inset
%plot. The inset illustrates that the asymptotic part of the relaxation follows
%the logarithmic law.}}{}{fig1.eps}{\special{ language "Scientific Word";
%type "GRAPHIC";  maintain-aspect-ratio TRUE;  display "FRAME";
%valid_file "F";  width 3.1012in;  height 3.1012in;  depth 0pt;
%original-width 7.8196in;  original-height 8.0601in;  cropleft "0";
%croptop "1.0312";  cropright "0.9760";  cropbottom "0.0843";
%filename 'Fig1.eps';file-properties "XNPEU";}}}%
%BeginExpansion
\begin{figure}
[ptb]
\begin{center}
\includegraphics[
trim=0.000000in 0.679466in 0.187670in -0.251476in,
height=3.1012in,
width=3.1012in
]%
{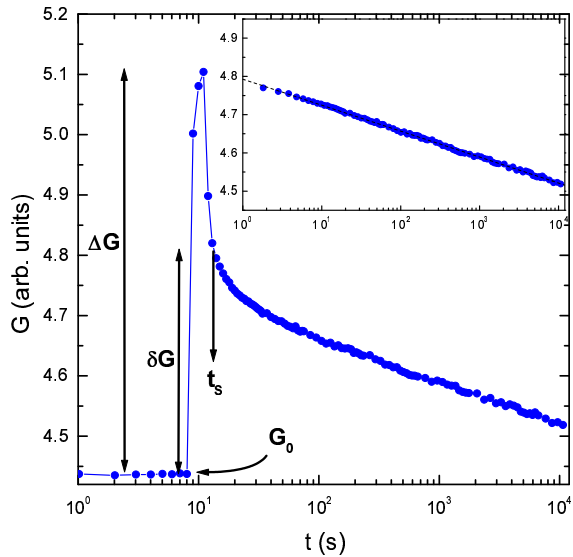}%
\caption{IR excitation protocol applied on a 600\AA ~In$_{x}$O film and
R$_{\square}$=50M$\Omega.$ The parameters that characterize the results are
marked by arrows; $\Delta$G is the conductance initial jump, $\delta$G is the
initial amplitude for the slowly relaxing part which is defined in this work
as the value of the excess conductance at t=t$_{S}$ which is the time
corresponding to t=1s of the inset plot. The inset illustrates that the
asymptotic part of the relaxation follows the logarithmic law.}%
\end{center}
\end{figure}
%EndExpansion

The qualitative features shown in figure 1 for the In$_{x}$O sample were
obtained on all electron-glass systems studied in this work (six different
materials). In terms of the balance between $\delta$G and $\Delta$G-$\delta$G
however, the results varied considerably among these systems. For example,
figure 2 shows IR-excitation results for two of the systems that yielded the
smallest slowly-relaxing conductance in response to the IR-protocol;
granular-aluminum and beryllium.%
%TCIMACRO{\FRAME{ftbpFU}{3.2335in}{3.0649in}{0pt}{\Qcb{Results of IR excitation
%for typical film of granular-aluminum with thickness 60\AA ~and R$_{\square}%
%$=48M$\Omega$ (a), and beryllium with R$_{\square}$=11M$\Omega$ and
%18\AA ~thick (this is one of the samples used in reference11).}}{}%
%{fig2.eps}{\special{ language "Scientific Word";  type "GRAPHIC";
%maintain-aspect-ratio TRUE;  display "FRAME";  valid_file "F";
%width 3.2335in;  height 3.0649in;  depth 0pt;  original-width 7.8196in;
%original-height 8.0601in;  cropleft "0";  croptop "0.9811";
%cropright "0.9650";  cropbottom "0.0944";
%filename 'Fig2.eps';file-properties "XNPEU";}}}%
%BeginExpansion
\begin{figure}
[ptb]
\begin{center}
\includegraphics[
trim=0.000000in 0.760873in 0.273686in 0.152336in,
height=3.0649in,
width=3.2335in
]%
{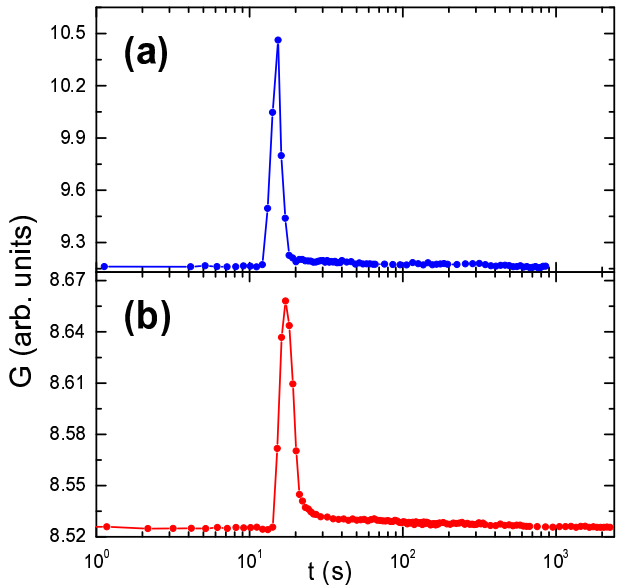}%
\caption{Results of IR excitation for typical film of granular-aluminum with
thickness 60\AA ~and R$_{\square}$=48M$\Omega$ (a), and beryllium with
R$_{\square}$=11M$\Omega$ and 18\AA ~thick (this is one of the samples used in
reference11).}%
\end{center}
\end{figure}
%EndExpansion

The very small $\delta$G/$\Delta$G exhibited by the sample in figure 2a was
reproduced on all our granular-aluminum samples independent of their
resistance (in the range 0.7-150M$\Omega$). In fact, in these samples the slow
component was so small as to raise the suspicion that the samples do not
exhibit glassy behavior at all! However, applying the F-protocol on the these
samples revealed the characteristic electron-glass features as shown in figure
3. Possible reasons for the poor response to the IR-excitation of granular
aluminum (in fact, of \textit{any} granular electron-glass) are discussed in
Section III D.
%TCIMACRO{\FRAME{ftbpFU}{3.1021in}{2.9395in}{0pt}{\Qcb{F-protocol using the
%same granular aluminum film as in figure 2a. Conductance is measured while the
%sample is under a fixed non-ohmic field F=20V/cm; note that G slowly increases
%as function of time (bottom graph). This is followed by a relaxation period
%where G is measured under ohmic field of F=2V/cm (upper graph, where the time
%is measured relative to the point where ohmic conditions were re-instated).
%This demonstrates that the process obeys the logarithmic relaxation law, a
%characteristic feature of electron-glasses.}}{}{fig3.eps}%
%maintain-aspect-ratio TRUE;  display "FRAME";  valid_file "F";
%width 3.1021in;  height 2.9395in;  depth 0pt;  original-width 7.8196in;
%original-height 8.0601in;  cropleft "0";  croptop "0.9815";
%cropright "0.9683";  cropbottom "0.0920";
%filename 'Fig3.eps';file-properties "XNPEU";}}}%
%BeginExpansion
\begin{figure}
[ptb]
\begin{center}
\includegraphics[
trim=0.000000in 0.741529in 0.247881in 0.149112in,
height=2.9395in,
width=3.1021in
]%
{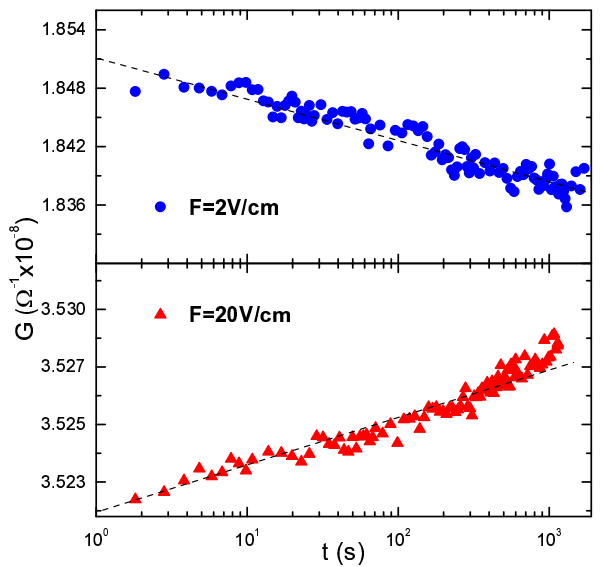}%
\caption{F-protocol using the same granular aluminum film as in figure 2a.
Conductance is measured while the sample is under a fixed non-ohmic field
F=20V/cm; note that G slowly increases as function of time (bottom graph).
This is followed by a relaxation period where G is measured under ohmic field
of F=2V/cm (upper graph, where the time is measured relative to the point
where ohmic conditions were re-instated). This demonstrates that the process
obeys the logarithmic relaxation law, a characteristic feature of
electron-glasses.}%
\end{center}
\end{figure}
%EndExpansion

Unless otherwise noted, the IR-protocol shown in this work involved a fixed
3-seconds exposure to the IR source. This is required to allow for a
meaningful comparison between different systems; all other things being equal,
longer exposures systematically yield larger values of $\delta$G/$\Delta$G. An
example is illustrated in figure 4. Note that during the time the IR is on,
and its intensity is held fixed, the conductance keeps going up. This
accumulated-over-time excess conductance translates into a larger $\delta$G
when the exposure is terminated. Essentially the same effect results by
increasing the bath temperature by $\Delta$T and holding it (see, e.g.,
reference [14]) for time $\tau$; the ensuing $\delta$G is then a monotonic
function of $\tau$ (as well as of $\Delta$T). Upon increasing the temperature,
the conductance promptly increases by $\Delta$G, but for a short $\tau,$
during which G(T+$\Delta$T) has not significantly changed, $\Delta$G would
decay as quickly as the original bath temperature is re-instated. The
3-seconds protocol is a compromise; it is short enough to minimize the heating
contribution to $\delta$G, and long enough with respect to the measurement
resolution-time (thus the recorded measurement of $\Delta$G is reasonably
accurate).%
%TCIMACRO{\FRAME{ftbpFU}{3.1012in}{2.821in}{0pt}{\Qcb{Results of IR excitations
%for a\ typical film of In$_{2}$O$_{3-x}$ (42\AA ~thick and R$_{\square}%
%$=10.5M$\Omega$), comparing the excess conductance $\Delta$G(t) ensuing after
%the source is turned off (after being on for 3s versus 180s).}}{}%
%{fig4.eps}{\special{ language "Scientific Word";  type "GRAPHIC";
%maintain-aspect-ratio TRUE;  display "FRAME";  valid_file "F";
%width 3.1012in;  height 2.821in;  depth 0pt;  original-width 7.8196in;
%original-height 8.0601in;  cropleft "0";  croptop "0.9803";
%cropright "0.9937";  cropbottom "0.1038";
%filename 'Fig4.eps';file-properties "XNPEU";}}}%
%BeginExpansion
\begin{figure}
[ptb]
\begin{center}
\includegraphics[
trim=0.000000in 0.836638in 0.049263in 0.158784in,
height=2.821in,
width=3.1012in
]%
{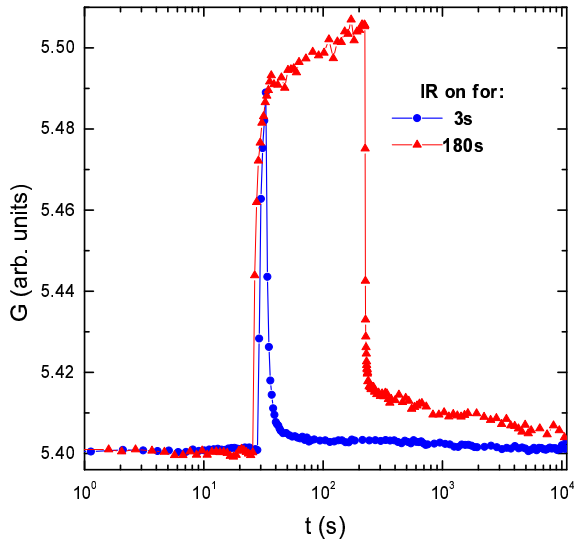}%
\caption{Results of IR excitations for a\ typical film of In$_{2}$O$_{3-x}$
(42\AA ~thick and R$_{\square}$=10.5M$\Omega$), comparing the excess
conductance $\Delta$G(t) ensuing after the source is turned off (after being
on for 3s versus 180s).}%
\end{center}
\end{figure}
%EndExpansion

Note that samples made of granular-aluminum, beryllium, and In$_{2}$O$_{3-x}$
yield $\delta$G/$\Delta$G values that are much smaller than the $\delta
$G/$\Delta$G obtained for the In$_{x}$O sample (figure 1). $\delta$G/$\Delta$G
is fairly constant in a given system and, in particular, it is not sensitive
to the value of the resistance; $\delta$G/$\Delta$G was found to be $\lesssim
$2\% in granular-aluminum (7 samples studied), 7-8\% in beryllium (5 samples),
8-12\% in In$_{2}$O$_{3-x}$ (more than 40 samples studied), and 40-55\% in
In$_{x}$O (22 samples with thicknesses 180-600\AA , with higher $\delta
$G/$\Delta$G values obtained for larger thicknesses). The ratio $\delta
$G/$\Delta$G will be taken as a measure of how efficient is the brief
IR-excitation process \cite{21}.

By our conjecture (c.f., Section I), a finite $\delta$G means that the charge
distribution in the system has been randomized to some degree. For that to
happen, at least \textit{some} absorption at the EM source frequency must take
place. To see how the variety in $\delta$G/$\Delta$G is reflected in the
optical properties of the samples we plot in figure 5 the optical transmission
of some of the actual films studied by the IR excitation.%
%TCIMACRO{\FRAME{ftbpFU}{3.147in}{2.879in}{0pt}{\Qcb{Optical transmission as
%function of wavelength for the materials that were measured by IR-excitation.
%Transmission through blank substrates were used as baseline in each case. Note
%the difference in transmission between the 200\AA ~and 600\AA ~In$_{x}$O
%samples indicating that the low transmission is mostly due to absorption
%(rather than reflection) throughout the entire range of wavelengths. The
%In$_{2}$O$_{3-x}$:Au and the In$_{2}$O$_{3-x}$:Pb samples both with 2 atomic
%\% doping levels.}}{}{fig5.eps}{\special{ language "Scientific Word";
%type "GRAPHIC";  maintain-aspect-ratio TRUE;  display "FRAME";
%valid_file "F";  width 3.147in;  height 2.879in;  depth 0pt;
%original-width 7.8196in;  original-height 8.0601in;  cropleft "0";
%croptop "0.9806";  cropright "1";  cropbottom "0.0936";
%filename 'Fig5.eps';file-properties "XNPEU";}}}%
%BeginExpansion
\begin{figure}
[ptb]
\begin{center}
\includegraphics[
trim=0.000000in 0.754425in 0.000000in 0.156366in,
height=2.879in,
width=3.147in
]%
{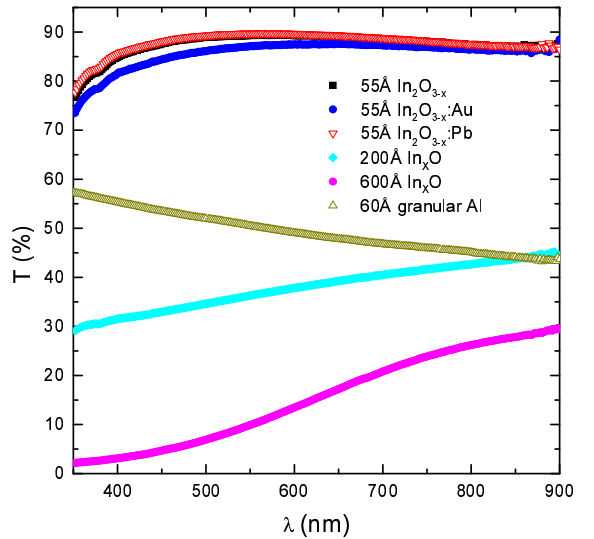}%
\caption{Optical transmission as function of wavelength for the materials that
were measured by IR-excitation. Transmission through blank substrates were
used as baseline in each case. Note the difference in transmission between the
200\AA ~and 600\AA ~In$_{x}$O samples indicating that the low transmission is
mostly due to absorption (rather than reflection) throughout the entire range
of wavelengths. The In$_{2}$O$_{3-x}$:Au and the In$_{2}$O$_{3-x}$:Pb samples
both with 2 atomic \% doping levels.}%
\end{center}
\end{figure}
%EndExpansion

At first sight, these data confirm the correlation between $\delta$G/$\Delta$G
and optical absorption. This is indeed the case for the samples for which
excitation data were shown above; the difference between In$_{2}$O$_{3-x}$ and
In$_{x}$O, and the thickness dependence in the latter are certainly in line
with this conjecture. The rather low value of the transmission of the granular
film is almost entirely due to reflection (and therefore the small $\delta
$G/$\Delta$G is a natural result of a weak absorption). This may also be the
case for beryllium (which was not measured due to lack of a suitable sample-size).

In addition to the materials discussed so far, figure 5 includes data on
In$_{2}$O$_{3-x}$ films doped with Au and Pb. Given that their transparency
shows only small changes relative to the undoped stuff (at least over the
range relevant for the IR source employed in these experiments), one might
expect that the IR-excitation results on these films should be similar to
those of the pure In$_{2}$O$_{3-x}$. It turns out that, in this case, a small
change makes a \textit{big} difference.

\subsection{ The case of In$_{2}$O$_{3-x}$:Au}

The original motivation to add gold to the In$_{2}$O$_{3-x}$ film was to test
the effect of spin-orbit scattering on the microwaves induced excess
conductance in the hopping regime. This actually yielded a null result but it
led to an intriguing response to IR radiation. Figure 6 shows the results of
the IR excitation protocol for a typical In$_{2}$O$_{3-x}$:Au sample (actually
the \textit{same} film depicted in figure 5 for optical transparency).%
%TCIMACRO{\FRAME{ftbpFU}{3.2439in}{2.9776in}{0pt}{\Qcb{The 3-second IR-protocol
%using a In$_{2}$O$_{3-x}$:Au sample (2 atomic \% Au doping, same sample as
%depicted in figure 5 for optical transmission), thickness is 55\AA ~and
%R$_{\square}$=11M$\Omega$. $\delta$G, and t$_{S}$ are as defined in figure 1.
%The inset illustrates the logarithmic relaxation law.}}{}{fig6.eps}%
%maintain-aspect-ratio TRUE;  display "FRAME";  valid_file "F";
%width 3.2439in;  height 2.9776in;  depth 0pt;  original-width 7.8196in;
%original-height 8.0601in;  cropleft "0";  croptop "0.9811";  cropright "1";
%cropbottom "0.0918";  filename 'Fig6.eps';file-properties "XNPEU";}}}%
%BeginExpansion
\begin{figure}
[ptb]
\begin{center}
\includegraphics[
trim=0.000000in 0.739917in 0.000000in 0.152336in,
height=2.9776in,
width=3.2439in
]%
{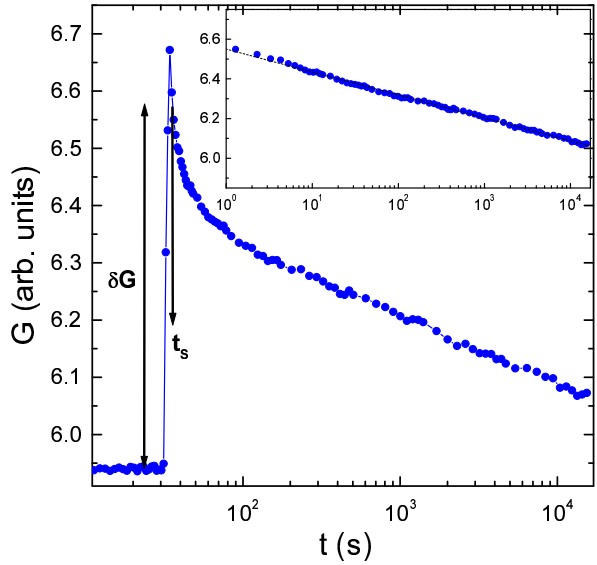}%
\caption{The 3-second IR-protocol using a In$_{2}$O$_{3-x}$:Au sample (2
atomic \% Au doping, same sample as depicted in figure 5 for optical
transmission), thickness is 55\AA ~and R$_{\square}$=11M$\Omega$. $\delta$G,
and t$_{S}$ are as defined in figure 1. The inset illustrates the logarithmic
relaxation law.}%
\end{center}
\end{figure}
%EndExpansion

Note that the slowly relaxing part of the excess conductance is not only much
larger than that of the pure material, it is also \textit{considerably larger}
than the 600\AA ~In$_{x}$O film despite its IR absorption being stronger
(compare figures 1 and 6). To check on the possibility that there is some
unusual absorption outside the range probed by the spectrometer (figure 5), we
tested the In$_{2}$O$_{3-x}$:Au sample in a FTIR extending the measurement to
1.3$\mu$m and compiled the results for this particular sample in figure 7. No
sign of an outstanding absorption feature could be found over the range
relevant for the used EM source.%
%TCIMACRO{\FRAME{ftbpFU}{3.4108in}{1.9527in}{0pt}{\Qcb{Optical transmission for
%one of the measured In$_{2}$O$_{3-x}$:Au (with 2 atomic \% Au) extended
%further into the infra-red regime by stitching data from a UV-visible
%spectrophotometer and FTIR. Both spectra were normalized by the measured
%transmission of the same blank substrate used for the sample deposition. The
%dashed lines delineate the spectral region where the IR source may have an
%appreciable intensity.}}{}{fig7.eps}{\special{ language "Scientific Word";
%type "GRAPHIC";  maintain-aspect-ratio TRUE;  display "FRAME";
%valid_file "F";  width 3.4108in;  height 1.9527in;  depth 0pt;
%original-width 7.8196in;  original-height 8.0601in;  cropleft "0";
%croptop "0.9792";  cropright "1";  cropbottom "0.4274";
%filename 'Fig7.eps';file-properties "XNPEU";}}}%
%BeginExpansion
\begin{figure}
[ptb]
\begin{center}
\includegraphics[
trim=0.000000in 3.444886in 0.000000in 0.167650in,
height=1.9527in,
width=3.4108in
]%
{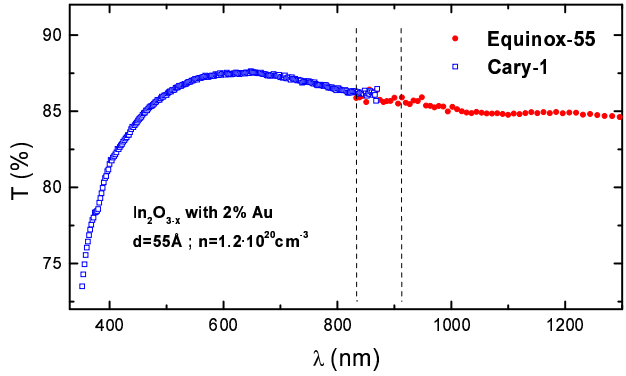}%
\caption{Optical transmission for one of the measured In$_{2}$O$_{3-x}$:Au
(with 2 atomic \% Au) extended further into the infra-red regime by stitching
data from a UV-visible spectrophotometer and FTIR. Both spectra were
normalized by the measured transmission of the same blank substrate used for
the sample deposition. The dashed lines delineate the spectral region where
the IR source may have an appreciable intensity.}%
\end{center}
\end{figure}
%EndExpansion

Large values of $\delta$G/$\Delta$G (in the range 70-90\%) were found in each
and every Au doped In$_{2}$O$_{3-x}$ studied in this work. These included
samples with R$_{\square}$ ranging between 250k$\Omega$ to 1.5G$\Omega$ , and
with doping levels of either 2 atomic \% (12 samples) or 4 atomic \% (5
samples). The latter group showed only slight advantage over the 2\% group in
terms of $\delta$G/$\Delta$G, which makes it interesting to measure the
dependence of $\delta$G/$\Delta$G on smaller doping levels. Unfortunately, our
current preparation and analysis methods are not yet accurate enough to
attempt such a study.

$\delta$G/$\Delta$G in a given sample depends to some degree on the
IR-excitation level as shown in figure 8. The decrease of $\delta$G/$\Delta$G
with IR intensity shown in the figure presumably reflects the saturation in
$\delta$G at these temperatures; the typical saturation value for $\delta$G by
gate excitation for a sample with the same R$_{\square}$ and measured under
similar conditions is $\approx$10\% (see, e.g., [11]) and probably also due to
the growing weight of the `heating' effect that accompanies the illumination.
At any rate, it is clear that the Au-doping has a dramatic influence on the
efficiency of IR-excitation process in the In$_{2}$O$_{3-x}$ samples. The
question is why does the addition of gold increase $\delta$G/G, or in other
words, why does it enhance the system randomization ?%
%TCIMACRO{\FRAME{ftbpFU}{3.2232in}{2.8504in}{0pt}{\Qcb{The dependence of the
%relative change of conductance $\Delta$G/G$_{0}$, and the slow-relaxation
%component $\delta$G/$\Delta$G (c.f., figure 1 for definition of G$_{0}$ and
%$\Delta$G), on the IR-source excitation current (each taken with the 3-second
%protocol). Results are for In$_{2}$O$_{3-x}$:Au (with 4 atomic \% Au),
%thickness 55\AA ~and R$_{\square}$=41M$\Omega$.}}{}{fig8.eps}%
%maintain-aspect-ratio TRUE;  display "FRAME";  valid_file "F";
%width 3.2232in;  height 2.8504in;  depth 0pt;  original-width 7.8196in;
%original-height 8.0601in;  cropleft "0";  croptop "0.9838";  cropright "1";
%cropbottom "0.1270";  filename 'Fig8.eps';file-properties "XNPEU";}}}%
%BeginExpansion
\begin{figure}
[ptb]
\begin{center}
\includegraphics[
trim=0.000000in 1.023633in 0.000000in 0.130574in,
height=2.8504in,
width=3.2232in
]%
{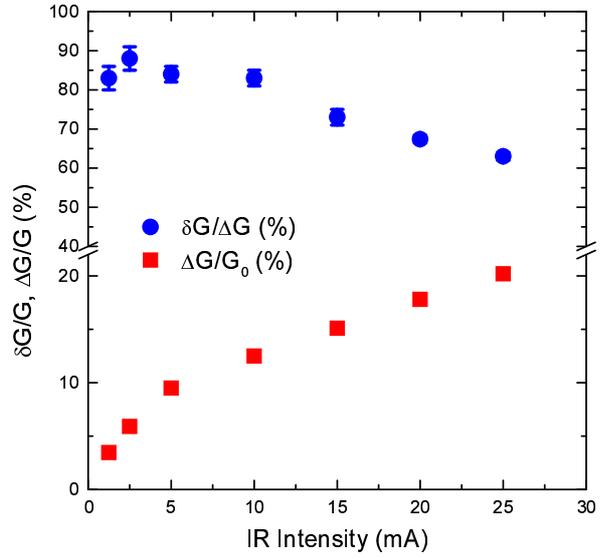}%
\caption{The dependence of the relative change of conductance $\Delta$%
G/G$_{0}$, and the slow-relaxation component $\delta$G/$\Delta$G (c.f., figure
1 for definition of G$_{0}$ and $\Delta$G), on the IR-source excitation
current (each taken with the 3-second protocol). Results are for In$_{2}%
$O$_{3-x}$:Au (with 4 atomic \% Au), thickness 55\AA ~and R$_{\square}%
$=41M$\Omega$.}%
\end{center}
\end{figure}
%EndExpansion

As argued above, randomizing the system entails an energy cost. We must
therefore accept that the addition of gold somehow enhances the absorption of
energy in the $\approx$0.88$\pm$0.5$\mu$m wavelength, and given that the
optical gap of In$_{2}$O$_{3-x}$ is at much higher energy (of order 3.7eV),
this must be an \textit{intra-}band absorption \cite{22}. In a clean metal
this process has an extremely small oscillator-strength due to
momentum-conservation constraints. Absorption becomes more likely when defects
or impurities are present. Therefore, it is natural to look for changes in the
microstructure of the In$_{2}$O$_{3-x}$:Au films relative to the undoped
material. Such changes are indeed observed in high resolution TEM micrographs
(figure 9). The micrograph for the In$_{2}$O$_{3-x}$:Au exposes a much higher
density of grain-boundaries and various structural defects than those present
in In$_{2}$O$_{3-x}$(figure 9a and 9b respectively). However, the defective
microstructure by itself cannot be the reason for the enhanced $\delta
$G/$\Delta$G. That can be seen by studying an \textit{undoped} In$_{2}%
$O$_{3-x}$ film that is as defective as In$_{2}$O$_{3-x}$:Au, which may be
prepared by crystallizing it at elevated temperatures. A TEM micrograph of
such film is shown in figure 9c (compare with figure 9a). A batch of these
tortured films were prepared and measured, and the 3-seconds IR protocol
performed on them yielded $\delta$G/$\Delta$G values of order 8-12\%, which
are quite similar to the results of the In$_{2}$O$_{3-x}$ films crystallized
at 525K (and have the structure as in figure 9b).%
%TCIMACRO{\FRAME{ftbpFU}{3.2975in}{3.9055in}{0pt}{\Qcb{TEM micrographs
%illustrating the microstructure of three versions of crystalline indium-oxide
%films all 55\AA ~thick and taken with identical magnification (the bar in the
%lower right corner of each micrograph is 100nm). (\textbf{a}) - In$_{2}%
%$O$_{3-x}$:Au (with 2 atomic \% Au) crystallized at 525K. (\textbf{b}) -
%In$_{2}$O$_{3-x}$ (\textit{undoped}) crystallized at 525K. (\textbf{c}) -
%In$_{2}$O$_{3-x}$ (\textit{undoped}) crystallized at 850K.}}{}{fig9.eps}%
%maintain-aspect-ratio TRUE;  display "FRAME";  valid_file "F";
%width 3.2975in;  height 3.9055in;  depth 0pt;  original-width 16.8024in;
%original-height 19.9218in;  cropleft "0";  croptop "1";  cropright "1";
%cropbottom "0";  filename 'Fig9.eps';file-properties "XNPEU";}}}%
%BeginExpansion
\begin{figure}
[ptb]
\begin{center}
\includegraphics[
height=3.9055in,
width=3.2975in
]%
{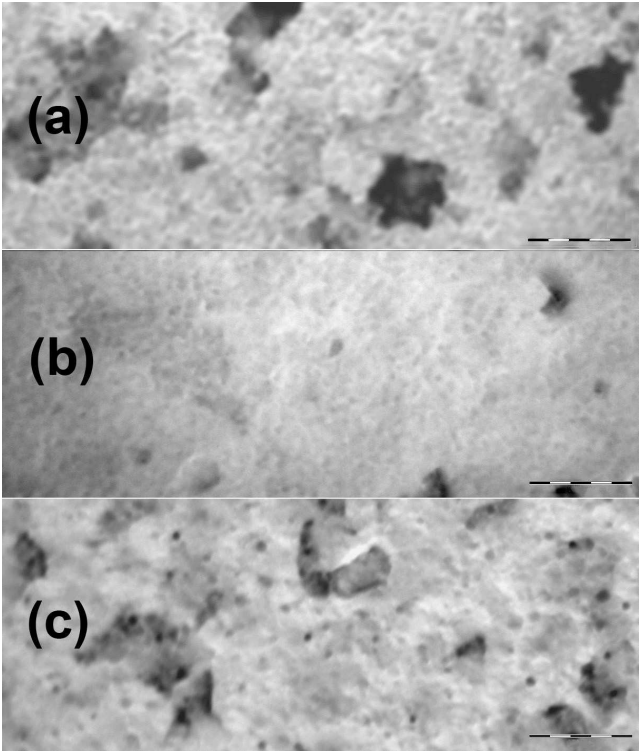}%
\caption{TEM micrographs illustrating the microstructure of three versions of
crystalline indium-oxide films all 55\AA ~thick and taken with identical
magnification (the bar in the lower right corner of each micrograph is 100nm).
(\textbf{a}) - In$_{2}$O$_{3-x}$:Au (with 2 atomic \% Au) crystallized at
525K. (\textbf{b}) - In$_{2}$O$_{3-x}$ (\textit{undoped}) crystallized at
525K. (\textbf{c}) - In$_{2}$O$_{3-x}$ (\textit{undoped}) crystallized at
850K.}%
\end{center}
\end{figure}
%EndExpansion

We conclude therefore that, if structural defects are responsible for the
$\delta$G enhancement, they must be of a more specific nature.

Another possibility that comes to mind is that it is the massiveness of the Au
impurities that plays a role. This conjecture is amenable to an experimental
test: replace gold with another heavy element. The choice we made was to use
Pb (that is actually heavier than Au) as the dopant. The results of these
attempts are described next.

\subsection{Not all that's heavy is (good as) Gold}

Three batches of In$_{2}$O$_{3-x}$:Pb films with doping levels of 2 atomic \%
were made and measured in this study with R$_{\square}$ (at T$\approxeq$4K)
ranging from 0.7M$\Omega$ to 250M$\Omega$. In terms of general transport
properties, such as current-voltage characteristics, the Pb-doped films behave
similarly to the Au-doped samples (for comparable resistance, sample-size,
etc.). The 2\%-doped Au and Pb systems also had similar carrier-concentration,
n=(1-2.2)$\cdot$10$^{20}$cm$^{-3}$ (measured via Hall effect at room
temperature), a somewhat higher value than that of the undoped indium-oxide,
which is usually in the range (4-8)$\cdot$10$^{19}$cm$^{-3}$. Finally, just
like the Au-doped samples, all In$_{2}$O$_{3-x}$:Pb samples that were in the
hopping regime showed electron-glass effects and exhibited slow relaxation
following IR-excitation. However, unlike the Au-doped samples, their IR
randomization efficiency turned out to be lower, and moreover, it was a
function of the sample age.

Typical results of the 3-second IR-excitation protocol are shown in figure 10.
These are results of measurements on the \textit{same} In$_{2}$O$_{3-x}$:Pb
sample but taken at different times; the data in figure 10a were taken a day
after preparation while those in figure 10b were taken after the sample was at
room temperatures for eleven months. The latter shows significantly smaller
$\delta$G/$\Delta$G. This complication was found when a sample that has been
previously measured was re-inserted into the cryostat for further measurements
after spending a couple of weeks at room temperatures. It then yielded
$\delta$G/$\Delta$G$\approx$17\%, already a smaller value than the one-day-old
$\delta$G/$\Delta$G$\approx$30\% result. It is remarkable that the changes
with time in the In$_{2}$O$_{3-x}$:Pb samples so well reflected in $\delta
$G/$\Delta$G, seem to be so subtle as to have escaped our structural and
analytical characterization. These included TEM, optical spectroscopy and XPS analysis.

The surprising In$_{2}$O$_{3-x}$:Pb results, prompted us to test the In$_{2}%
$O$_{3-x}$:Au samples over a similar period of time. However, unlike the
In$_{2}$O$_{3-x}$:Pb samples, the results of $\delta$G$/\Delta$G for In$_{2}%
$O$_{3-x}$:Au samples were independent of their age. In addition to the
difference between the two systems with regard to sample age, the maximum
value of $\delta$G/$\Delta$G we obtained to date with `young' In$_{2}$%
O$_{3-x}$:Pb was of the order of 35\%, still considerably less than in the
gold-doped system. The source of these differences appears to be related to
chemistry which dictates the way the Au and Pb are incorporated into the
In$_{2}$O$_{3-x}$ lattice. Indeed, XPS studies showed a clear difference
between In$_{2}$O$_{3-x}$:Au and In$_{2}$O$_{3-x}$:Pb.
%TCIMACRO{\FRAME{ftbpFU}{3.2162in}{2.9551in}{0pt}{\Qcb{Results for the 3-second
%IR-protocol using In$_{2}$O$_{3-x}$:Pb sample (with 2 atomic \% Pb doping),
%and thickness 55\AA . (\textbf{a}) 24 hours after preparation, R$_{\square}%
%$=8.9M$\Omega$. (\textbf{b}) 11 months after preparation, R$_{\square}%
%$=7.7M$\Omega$. Note the decrease in $\delta$G/$\Delta$G of the old sample.}%
%}{}{fig10.eps}{\special{ language "Scientific Word";  type "GRAPHIC";
%maintain-aspect-ratio TRUE;  display "FRAME";  valid_file "F";
%width 3.2162in;  height 2.9551in;  depth 0pt;  original-width 7.8196in;
%original-height 8.0601in;  cropleft "0";  croptop "0.9772";  cropright "1";
%cropbottom "0.0870";  filename 'Fig10.eps';file-properties "XNPEU";}}}%
%BeginExpansion
\begin{figure}
[ptb]
\begin{center}
\includegraphics[
trim=0.000000in 0.701229in 0.000000in 0.183770in,
height=2.9551in,
width=3.2162in
]%
{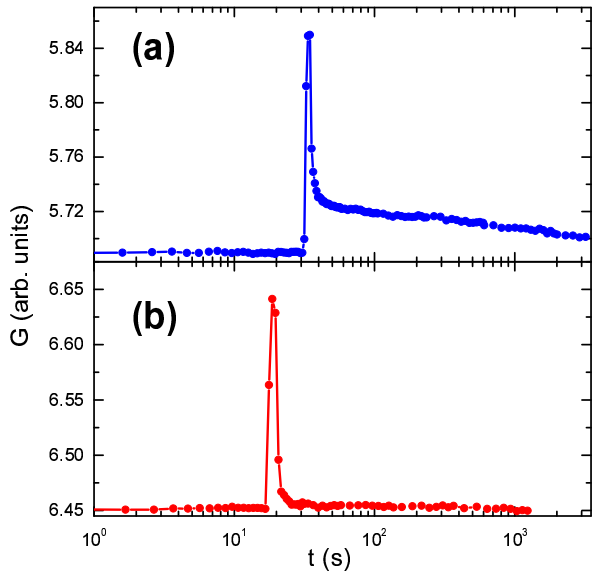}%
\caption{Results for the 3-second IR-protocol using In$_{2}$O$_{3-x}$:Pb
sample (with 2 atomic \% Pb doping), and thickness 55\AA . (\textbf{a}) 24
hours after preparation, R$_{\square}$=8.9M$\Omega$. (\textbf{b}) 11 months
after preparation, R$_{\square}$=7.7M$\Omega$. Note the decrease in $\delta
$G/$\Delta$G of the old sample.}%
\end{center}
\end{figure}
%EndExpansion

\texttt{ }Figure 11 compares some of the characteristic binding energies for
Au and Pb obtained from XPS scans of the respective samples where these
elements were incorporated. The binding energies of the Pb 4f levels are
consistent with Pb$_{3}$O$_{4}$ suggesting that the Pb resides in the In$_{2}%
$O$_{3-x}$ matrix by forming Pb$_{2}$O and PbO, (i.e., with oxidation state of
either +2 or +4). Gold, on the other hand, retains its neutral valency. There
is a small upward shift of the binding energies of the Au:4f electrons by
$\approx$0.1eV, presumably due to the reduced screening associated with the
In$_{2}$O$_{3-x}$ matrix relative to the reference (metallic) Au film.
Electron diffraction patterns taken of samples with level doping smaller than
6-8\% reveal just the bcc rings of In$_{2}$O$_{3}.$ No evidence for Au
clusters is detected by either the diffraction patterns or in high resolution
imaging of these films. This is consistent with previous studies \cite{23,24}
of In$_{2}$O$_{3-x}$:Au where it was concluded that the Au atoms reside in
oxygen vacancies (or possibly di-vacancies).%
%TCIMACRO{\FRAME{ftbpFU}{3.1678in}{3.0217in}{0pt}{\Qcb{Binding energies for the
%4f states of Au and Pb. XPS scans performed on the actual samples used for the
%IR-excitation study. The Au reference is an evaporated film of gold on a glass
%substrate.}}{}{fig11.eps}{\special{ language "Scientific Word";
%type "GRAPHIC";  maintain-aspect-ratio TRUE;  display "FRAME";
%valid_file "F";  width 3.1678in;  height 3.0217in;  depth 0pt;
%original-width 7.8196in;  original-height 8.0601in;  cropleft "0";
%croptop "0.9871";  cropright "1";  cropbottom "0.0618";
%filename 'Fig11.eps';file-properties "XNPEU";}}}%
%BeginExpansion
\begin{figure}
[ptb]
\begin{center}
\includegraphics[
trim=0.000000in 0.498114in 0.000000in 0.103975in,
height=3.0217in,
width=3.1678in
]%
{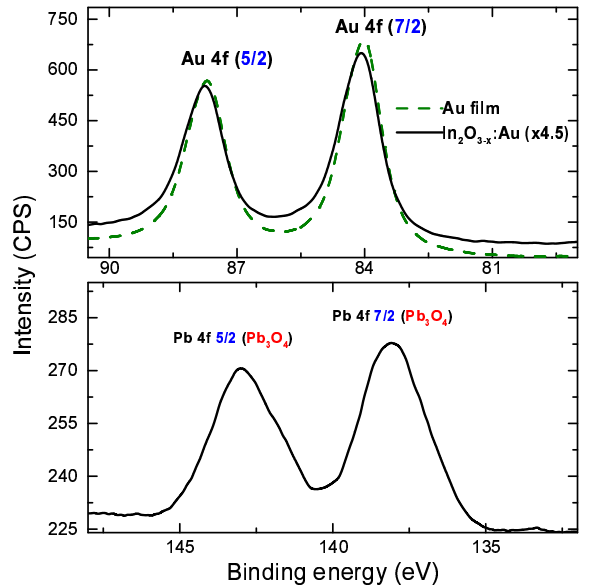}%
\caption{Binding energies for the 4f states of Au and Pb. XPS scans performed
on the actual samples used for the IR-excitation study. The Au reference is an
evaporated film of gold on a glass substrate.}%
\end{center}
\end{figure}
%EndExpansion

Being chemically bonded with oxygen the Pb atoms reside in the In$_{2}%
$O$_{3-x}$ as either, interstitials, or they replace indium atoms in the
In$_{2}$O$_{3-x}$ lattice. In the later case they play a similar role as Sn,
which is a common dopant in indium-oxide. This is presumably the cause of the
higher carrier-concentration in In$_{2}$O$_{3-x}$:Pb samples relative to the
undoped material.

That the Pb and Au atoms occupy different sites in the indium-oxide lattice is
consistent with the quite different spin-orbit scattering contributed by these
dopants in indium-oxides. Gold-doped samples were used in several past
experiments in both the hopping \cite{23} and the diffusive \cite{24} regimes,
and in both cases the incorporation of Au was shown to enhance spin-orbit
scattering. Lead-doping on the other hand, did not seem to be effective in
this regard, which may seem surprising considering that Pb is a heavy
impurity. The reason for that is that the main source of elastic scattering in
In$_{2}$O$_{3-x}$ is oxygen vacancies, not the heavy In atoms (or the Pb
residing on In sites). Effective spin-orbit scattering occurs only when a
heavy atom like Au resides on an oxygen site. A similar situation occurs in
n-type GaAs; spin orbit scattering is weak despite the presence of the heavy
Ga. In both cases it is the large affinity of electrons to (the light) anion
that determines the relative importance of spin-orbit scattering \cite{25}.

Note that the weak spin-orbit scattering of In$_{2}$O$_{3-x}$ and In$_{2}%
$O$_{3-x}$:Pb is also a property of beryllium and aluminum, being light
elements. It is intriguing that these materials also share a much smaller IR
excitation efficiency relative to In$_{2}$O$_{3-x}$:Au where spin-orbit
scattering is strong. Spin-orbit scattering has been known to alter optical
transitions probabilities e.g., by lifting-off selection rules \cite{26}.
However, attempts to find an explanation for the enhancement effect of the Au
doping along this line have not produced anything conclusive.

The different positions where the Au and Pb dopants reside in the In$_{2}%
$O$_{3-x}$ lattice should also have consequences for the band structure of the
material. Band structure calculations suggest that the bottom of the
conduction band of indium-oxide is mainly formed from indium 5s-states
hybridized with oxygen 2s-states \cite{27}. This could be the reason for the
modified carrier-concentration due to the Au inclusion. More generally, for
doping levels of few percents there is probably an impurity band associated
with the foreign element, and the only question is at what energy range it
forms; if it overlaps with any of the host relevant bands, a profound change
in the material properties may take place.

Modifications in the nature of the conduction-band states is of vital
importance to the oscillator-strength for the intraband transitions, which in
turn, determine the efficiency of the IR excitation. Such changes are actually
observable in the optical spectra of the In$_{2}$O$_{3-x}$:Au and the In$_{2}%
$O$_{3-x}$:Pb systems. These may have been easy to ignore in figure 5 because
they are relatively weak (especially at the energy range where the IR
absorption may occur). To resolve better the more subtle details we compare in
figure 12 the ratio between the transmission versus wavelength of the undoped
In$_{2}$O$_{3-x}$relative to the two systems under consideration. Note that
the spectra is now extended somewhat further into the near UV where the
relative changes in the transmission stand out more clearly.%
%TCIMACRO{\FRAME{ftbpFU}{3.1116in}{2.7899in}{0pt}{\Qcb{Transmission of the 2
%atomic \% doped Au and Pb films relative to that of pure In$_{2}$O$_{3-x}$
%film of the same thickness (55\AA ), as function of the wavelength. All three
%films were deposited on quartz substrates to allow for better resolution in
%the near UV.}}{}{fig12.eps}{\special{ language "Scientific Word";
%type "GRAPHIC";  maintain-aspect-ratio TRUE;  display "FRAME";
%valid_file "F";  width 3.1116in;  height 2.7899in;  depth 0pt;
%original-width 7.8196in;  original-height 8.0601in;  cropleft "0";
%croptop "0.9791";  cropright "1";  cropbottom "0.1105";
%filename 'Fig12.eps';file-properties "XNPEU";}}}%
%BeginExpansion
\begin{figure}
[ptb]
\begin{center}
\includegraphics[
trim=0.000000in 0.890641in 0.000000in 0.168456in,
height=2.7899in,
width=3.1116in
]%
{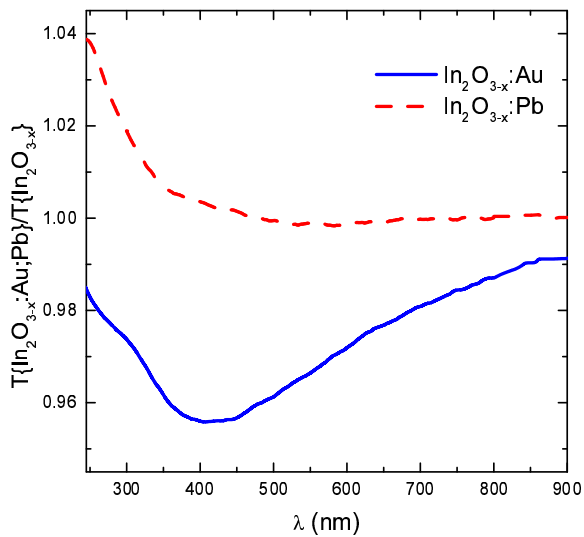}%
\caption{Transmission of the 2 atomic \% doped Au and Pb films relative to
that of pure In$_{2}$O$_{3-x}$ film of the same thickness (55\AA ), as
function of the wavelength. All three films were deposited on quartz
substrates to allow for better resolution in the near UV.}%
\end{center}
\end{figure}
%EndExpansion

Focusing first on the In$_{2}$O$_{3-x}$:Pb spectrum one notes that over most
of the range relevant for the IR-excitation there is no evidence for
absorption over that of the pure In$_{2}$O$_{3-x}$ and therefore no reason why
this sample should exhibit different excitation behavior. There seems however
to be a higher threshold for the interband absorption in the Pb-doped sample.
This is due to the Burstein shift associated with the higher
carrier-concentration n. A similar effect has been observed in undoped
In$_{2}$O$_{3-x}$ where n was increased by UV-treatment \cite{28}, as well as
in many studies of indium-oxides doped with Sn \cite{29}.

The spectrum for the In$_{2}$O$_{3-x}$:Au sample does indicate excess
absorption throughout the entire studied range, with a broad $\approxeq$4\%
peak around $\lambda\approx$400nm. A similarly shaped feature, at a lower
energy, was reported in quantum dots and ascribed to an intraband absorption
\cite{30}. The spectroscopic features at wavelengths $\eqslantless$800nm are
not directly relevant for the excitation process (as the IR source has
negligible contribution at this range) but they may be indicative of changes
in the band structure. The nature of the modifications in the electronic
states that give rise to the enhanced intraband absorption are yet to be determined.

\subsection{The routes for system randomization}

Once intraband transitions are turned on, system randomization becomes
possible (the $\hbar\omega$
%TCIMACRO{\TEXTsymbol{>}}%
%BeginExpansion
$>$%
%EndExpansion
E$_{C}$ condition is easily fulfilled as E$_{C}$ in In$_{2}$O$_{3-x}$ is
$\lesssim$10meV) and may proceed along two main routes. Either route will
drive the system far from equilibrium, achieving a long-lived, enhanced
conductance state. The first is a direct transition from an equilibrium site
at r$_{i}$ to another localized state at r$_{j}$, where a photon is absorbed
to make up for the difference in energy. Alternatively, excitation occurs by a
two-stage process where a vertical on-site excitation decays via phonon
emission, which in turn supply the energy for the r$_{i}\rightarrow$r$_{j}$ transition.

For an effective randomization process, many such transitions should take
place during the excitation and, in addition to energy conservation, this
requires reasonably good overlap of the wavefunctions involved in the
transitions. It is here that the Anderson insulator has a distinct advantage
over a granular system. The determining factor in the granular system is the
grain-size which has a fixed value set by the system microstructure. The
equivalent of a "grain" in the Anderson insulator is the localization length
$\xi$ which is determined by multiple scattering off the disorder potential,
and importantly, it is a function of energy. Realistic low temperature studies
of Anderson insulators rarely if ever involve samples where the Fermi energy
E$_{F}$ is far from the mobility edge (or, in the two-dimensional case, an
energy region where $\xi$ is not the limiting factor for the overlap). Optical
excitation energy may be large enough to probe states with essentially
extended states. The Fermi energy of In$_{2}$O$_{3-x}$ for example, is of
order 0.1eV, and $\hbar\omega\simeq$1eV should be sufficiently large to secure
significant overlap of most r$_{i}\rightarrow$r$_{j}$ transitions. In a
granular system only special pairs of initial-final states (possibly related
by Mahan exciton conditions \cite{31}) might be available for favorable
r$_{i}\rightarrow$r$_{j}$ transitions. Moreover, when the grains are made of
the pure element (as is the case of granular aluminum), the probability of
intraband absorption is very small to start with even though the spatial
overlap for \textit{on-site} transitions may be excellent.

It should still be possible to drive a granular system far from equilibrium if
the EM\ frequency is high enough to cause photoemission. Electrons ejected
from the sample will be replaced by the neutralizing charge that will flow
into the system from the contacts, which in turn should effectively randomize
the system in a similar way that changing a gate voltage would have \cite{32}.
The photon energy required to achieve these conditions for a typical granular
system is probably in the vacuum-UV range, which may be large enough to induce
structural changes in the sample. This complication could however be
monitored, and taken into account if use is made of mesoscopic samples where
effects associated with the potential-landscape set by the ions were shown to
be separable from the electron-glass dynamics \cite{33}.

\subsection{Summary}

We have studied in this work several aspects of electron-glass excitation by
near IR radiation. Six different materials were used in this study, all
showing similar electron-glass features when driven far from equilibrium. In
terms of their response to the IR excitation however, they exhibit widely
different efficiency defined by $\delta$G/$\Delta$G. In gold doped
indium-oxide this measure of efficiency could be as large as 90\% while the
best figure we obtained on granular-aluminum samples never exceeded 2\%. In
general, the efficiency of taking the system far from equilibrium by IR
radiation appears to be low, the case of In$_{2}$O$_{3-x}$:Au being an
exception (and to a degree, also In$_{x}$O). Based on the phenomenology
presented here it seems obvious that the main reason for the poor response of
most systems is not energy conservation but rather some selection-rules that
suppress optical transitions. In the case of In$_{2}$O$_{3-x}$ the relevant
transitions for the $\hbar\omega$ used here involve the conduction band
electronic states (intraband transitions). The addition of Au atoms to the
lattice apparently relaxes these selection-rules leading to real absorption
that in turn randomizes the system. A weaker effect occurs when Pb atoms are
incorporated into the structure, and moreover, the enhanced $\delta$G/$\Delta
$G fades away with time. Chemical analysis indicate that the Pb atoms occupy
different sites in the In$_{2}$O$_{3-x}$ lattice than the Au atoms, which may
provide the clue for the difference in their effects on the IR-excitation
efficiency. Microscopic band structure calculations are needed to elucidate
these issues.

It is remarkable that the enhancement of $\delta$G/$\Delta$G produced by the
Au doping (and to some degree, by the Pb doping as well), is much more
conspicuous than the respective effects observable in any of the spectroscopic
and analytical tools used here. This sensitivity of the electron-glass may
prove useful in studies of basic condensed matter phenomena.

Driving electron-glasses far from equilibrium by high-frequency EM radiation
has the advantages and flexibility of modern light-sources. Having an
efficient system like the In$_{2}$O$_{3-x}$:Au\texttt{ }opens possibilities of
performing experiments that hitherto were not feasible. An examples for such
an experiment is spatially localized excitation, which could yield unique
information pertaining to the nature of the relaxation in electron-glasses.
Also, excitation studies using sources with different photon energies should
help to identify the energy scales relevant to the studied phenomena.

Illuminating discussions with O. Agam, and N. Kaplan are gratefully
acknowledged. This work greatly benefited from many helpful discussions with
M. Pollak, B. Shapiro, and other participants of the Electron-Glasses program
held at Santa-Barbara. The author expresses gratitude for the hospitality and
support of the KAVLI institute that hosted this program. Thanks are due to X.
M. Xiong, and P. W. Adams for the Be samples. This research was supported by a
grant administered by the US Israel Binational Science Foundation and by the
Israeli Foundation for Sciences and Humanities.

\texttt{ }


\begin{thebibliography}{99}
\bibitem {21}The normalization by $\Delta$G takes care (at least partially) of
the variation in the illumination level from sample to sample, which may have
different sizes, and not being placed at the exact position relative to the IR-beam.
\end{thebibliography}
\end{document}